\documentstyle[12pt]{article}
\begin{document}
\begin{center}
\vfill
\large\bf{Supersymmetry on $AdS_3$ and $AdS_4$}\\
\end{center}
\vfill
\begin{center}
D.G.C. McKeon\\
Department of Applied Mathematics\\
University of Western Ontario\\
London, Ontario, Canada\\
N6A 1B7\\
\vspace{2cm}
C. Schubert\\
Department of Physics and Geology\\
University of Texas Pan American\\
1201 West University Drive\\
Edinburg, Texas   78541-2999\\
U.S.A.
\vspace{1 cm}

Published in \\
{\sl ÊClassical and Quantum Gravity, volume 21, issue 13, 3337-3345}\\
\end{center}

\vspace{1cm}
\section{Abstract}

We consider a supersymmetric extension of the algebra associated with three and
four dimensional anti de Sitter space. A representation of the
supersymmetry operators in an embedding superspace is given. Supersymmetric invariant
models are constructed in the embedding superspace associate with $AdS_3$. 

\section{Introduction}
{\sf The supersymmetric extensions of the symmetry groups associated with
flat spaces have been closely analyzed. The supersymmetry algebras that
arise from spaces of constant curvature have also been considered [1],
 and superfield models have been constructed for them [7]. These superspace models in ref. [7] are formulated on the curved spaces themselves. Recently, 
a representation of the SUSY operators in superspace connected 
with $AdS_2$ and $S_2$ was found [2], using a different approach 
based on embedding coordinates (i.e., three dimensional flat space coordinates), akin to the one taken by Dirac 
for treating fields in four dimensional de Sitter space. In the 
present paper, the SUSY algebras arising from $AdS_3$ and $AdS_4$ 
are examined along similar lines.}
For $AdS_3$, a superspace whose Bosonic
coordinates constitute a four dimensional ($2 + 2D$) space with metric
$\eta^{\mu\nu} = {\rm{diag}}(+ + - -)$ in which $AdS_3$ is embedded is
devised. This also involves using Grassmann coordinates which are four component Majorana-Weyl spinors defined in $(2 + 2)$ dimensions. The supersymmetry operators are given a representation in this
embedding superspace. The ability to do so depends on the (anti-) self duality of the
spin tensor for rotations.  Superfield actions are written in this
superspace and then re-expressed using component fields.  A
representation in superspace of the supersymmetry operators associated
with $AdS_4$ is similarly devised, though a superfield action in this embedding
superspace does not appear to be immediately feasable.  This is in contrast to ref. [7] in which a generalization of $AdS_4$ itself to superspace using the coset $OSp(1,4)/O(1,3)$ is used to formulate superspace models.

An analysis of the positive energy unitary representations of the $AdS_3$
superalgebras appears in ref. [3].

\section{$AdS_3$ Superspace}

The Bosonic space we consider is a three dimensional surface defined by
the equation
$$x^2 = \eta_{\mu\nu} x^\mu x^\nu \equiv (x^0)^2 + (x^1)^2 - (x^2)^2 -
(x^3)^2 = a^2 .\eqno(1)$$
The Dirac matrices associated with the $(2+2)$ dimensional embedding space are taken to be real:
$$\gamma^\mu = \left(\begin{array}{cc}
0 & \lambda^\mu\\
\overline{\lambda}^\mu & 0\end{array}\right)\eqno(2a)$$
with
$$\lambda^\mu = \left(1, i\tau^2, \tau^1, \tau^3\right)\eqno(2b)$$
$$\overline{\lambda}^\mu = \left(1, -i\tau^2, -\tau^1, -
\tau^3\right)\eqno(2c)$$
where $\tau^i$ is a Pauli spin matrix. These matrices satisfy
$$\lambda^\mu \overline{\lambda}^\nu + 
\lambda^\nu\overline{\lambda}^\mu = 2\eta^{\mu\nu} .\eqno(3)$$
If now we define
$$\sigma^{\mu\nu} = -\frac{1}{4} \left[\lambda^\mu\overline{\lambda}^\nu
- \lambda^\nu\overline{\lambda}^\mu\right]\eqno(4a)$$
$$\overline{\sigma}^{\mu\nu} = -\frac{1}{4} \left[\overline{\lambda}^\mu\lambda^\nu
- \overline{\lambda}^\nu\lambda^\mu\right]\eqno(4b)$$
then we find that
$$\left[\sigma^{\mu\nu},\sigma^{\lambda\sigma}\right] =
\eta^{\mu\lambda}\sigma^{\nu\sigma} -
\eta^{\nu\lambda}\sigma^{\mu\sigma} +
\eta^{\nu\sigma}\sigma^{\mu\lambda} -
\eta^{\mu\sigma}\sigma^{\nu\lambda}\eqno(5a)$$
$$\left[\overline{\sigma}^{\mu\nu},\overline{\sigma}^{\lambda\sigma}\right] =
\eta^{\mu\lambda}\overline{\sigma}^{\nu\sigma} -
\eta^{\nu\lambda}\overline{\sigma}^{\mu\sigma} +
\eta^{\nu\sigma}\overline{\sigma}^{\mu\lambda} -
\eta^{\mu\sigma}\overline{\sigma}^{\nu\lambda}.\eqno(5b)$$
Furthermore, there are the duality relations
$$\sigma^{\mu\nu} = -\frac{1}{2} \epsilon^{\mu\nu\lambda\sigma}\sigma_{\lambda\sigma}
\eqno(6a)$$
$$\overline{\sigma}^{\mu\nu} =+\frac{1}{2} \epsilon^{\mu\nu\lambda\sigma}
\overline{\sigma}_{\lambda\sigma}
\eqno(6b)$$
where $\epsilon^{0123} = \epsilon_{0123} = + 1$. Other useful relations are
$$\lambda^\mu\overline{\lambda}^\nu\lambda^\sigma = \eta^{\mu\nu}\lambda^{\sigma} -
\eta^{\mu\sigma}\lambda^\nu + \eta^{\nu\sigma}\lambda^\mu 
+\epsilon^{\mu\nu\sigma\rho}\lambda_\rho\eqno(7)$$
$$\tau^2 \lambda^\mu\tau^2 = \overline{\lambda}^{\mu T}\eqno(8a)$$
$$\tau^2 \sigma^{\mu\nu}\tau^2 = -\sigma^{\mu\nu T}.\eqno(8b)$$

Transformations that leave the $AdS_3$ space corresponding to the surface defined by eq. (1) invariant are generated
by
$$J^{\mu\nu} = -x^\mu\partial^\nu + x^\nu\partial^\mu .\eqno(9)$$
These satisfy the algebra
$$\left[J^{\mu\nu}, J^{\lambda\sigma}\right] = 
\eta^{\mu\lambda}J^{\nu\sigma} -
\eta^{\nu\lambda}J^{\mu\sigma} +
\eta^{\nu\sigma}J^{\mu\lambda} -
\eta^{\mu\sigma}J^{\nu\lambda}.\eqno(10)$$
If we now define
$$K^{\mu\nu} = \frac{1}{2}\left(J^{\mu\nu} - \frac{1}{2} \epsilon^{\mu\nu\lambda\sigma}
J_{\lambda\sigma}\right) = -\frac{1}{2} \epsilon^{\mu\nu\lambda\sigma}K_{\lambda\sigma}
\eqno(11a)$$
$$\overline{K}^{\mu\nu} = \frac{1}{2}\left(J^{\mu\nu} + \frac{1}{2} \epsilon^{\mu\nu\lambda\sigma}
J_{\lambda\sigma}\right) = +\frac{1}{2} \epsilon^{\mu\nu\lambda\sigma}\overline{K}_{\lambda\sigma}
\eqno(11b)$$
then both $K^{\mu\nu}$ and $\overline{K}^{\mu\nu}$ satisfy relations of the form of eq. (10).
This corresponds to the decomposition $SO(2,2) = SO(2,1) \times SO(2,1)$ [3].

Spinors in $2 + 2$ dimensions can be simultaneously chiral and Majorana [4,5]. The Majorana
condition is 
$$\psi = \psi_C \equiv C\overline{\psi}^T\eqno(12)$$
where
$$\overline{\psi} = \psi^\dagger A\eqno(13)$$
with
$$\gamma^{\mu \dagger} = A\gamma^\mu A^{-1}\eqno(14a)$$
$$\gamma^{\mu T} = -C\gamma^\mu C^{-1}.\eqno(14b)$$
With eq. (14), we have $A = -C = -i \gamma^2\gamma^3$ so that (12) becomes
$$\psi = \psi_C = \psi^* \eqno(15)$$
and (13) gives
$$\overline{\psi} = \psi^\dagger\left(\begin{array}{cc}
\tau^2 & 0\\
0 & \tau^2\end{array}\right).\eqno(16)$$

If $Q$ is a supersymmetry generator, taken to be a Majorana-Weyl
spinor in $(2+2)$ dimensions, then a suitable extension of the $AdS_3$ algebra of eq. (10) is
$$\left\lbrace Q, \tilde{Q}\right\rbrace = 2\sigma^{\mu\nu}
J_{\mu\nu}\eqno(17a)$$
$$\left[ J^{\mu\nu}, Q\right] = -\sigma^{\mu\nu}Q.\eqno(17b)$$
Here, the operators $J_{\mu\nu}$ are isometry generators on the four dimensional embedding space; in ref. [7] this has been decomposed into generators that live on the three dimensional surface of eq. (1).

In (17a), we have taken $\tilde{Q} \equiv Q^T\tau^2$ on account of (15)
and (16).  The algebra of (10), (17) satisfies the Jacobi identities;
proving this entails using the Fierz identity [2] for\\
$\Sigma^{\mu\nu} = \left(\begin{array}{cc}
\sigma^{\mu\nu} & 0\\
0 & \overline{\sigma}^{\mu\nu}\end{array}\right)$,
$$\displaystyle{\left(\Sigma^{\mu\nu}\right)_{ij}\left(\Sigma_{\mu\nu}\right)_{k\ell} =
-\frac{1}{2}\left(\Sigma^{\mu\nu}\right)_{i\ell}\left(\Sigma_{\mu\nu}\right)_{kj}
- \frac{3}{4}\left[\delta_{i\ell}\delta_{kj} +
\gamma_{i\ell}^5\gamma_{kj}^5\right]}\eqno(18)$$
with $\gamma^5 = -\gamma^0\gamma^1\gamma^2\gamma^3$. (One could also
have used $\overline{\sigma}^{\mu\nu}$ in place of $\sigma^{\mu\nu}$ in
(17).) 

In order to have a representation of the algebra of (10), (17) in this
superspace which is formulated as a generalization of the surface in $2 + 2$ dimensional space defined by eq. (1), we consider
$$Q_i = \left(\lambda^\mu\right)_{ij} \left(\partial_\mu\theta_j +
x_\mu\frac{\partial}{\partial\tilde{\theta}_j}\right)\eqno(19)$$
where
$$\frac{\partial}{\partial\theta_i} =
\frac{\partial\tilde{\theta}_j}{\partial\theta_i}
\frac{\partial}{\partial\tilde{\theta}_j} = \tau^2_{ij}
\frac{\partial}{\partial\tilde{\theta}_j}\;.\eqno(20)$$
From (19), it is apparent that (using (8a))
$$\tilde{Q} = \left(\tilde{\theta} \partial_\mu -
\frac{\partial}{\partial\theta}
x_\mu\right)\overline{\lambda}^\mu\eqno(21)$$
(We have introduced a Fermionic coordinate $\theta$ which is
a Majorana-Weyl Grassmann spinor in $2 + 2$ dimensions.)

It is apparent from (19) and (21) that
$$\left\lbrace Q_i, \tilde{Q}_j\right\rbrace =
2\left(\sigma^{\mu\nu}\right)_{ij} J_{\mu\nu}\eqno(22)$$
on account of the Fierz identity
$$\left(\lambda^\mu\right)_{ij}\left(\overline{\lambda}_\mu\right)_{k\ell}
= 2\delta_{i\ell}\delta_{kj}.\eqno(23)$$
Furthermore, by using eq. (7), we find that
$$\left[K^{\mu\nu}, Q_i\right] = -
\left(\sigma^{\mu\nu}\right)_{ij}Q_j .\eqno(24)$$
Note that (24), unlike (17b), is consistent with (6a) and (11a).

It is now possible to use the relation (6a) to rewrite (22) as
$$\left\lbrace Q_i, \tilde{Q}_j\right\rbrace =
2\left(\sigma^{\mu\nu}\right)_{ij}K_{\mu\nu}.\eqno(25)$$
Together, (24) and (25) constitute a supersymmetric extension of the
algebra associated with $AdS_3$.  One could also take
$$R_i = \overline{\lambda}_{ij}^\mu \left(\partial_\mu \theta_j +
x_\mu\frac{\partial}{\partial\tilde{\theta}_j}\right)\eqno(26)$$
in place of (19).  In this case we would have the superalgebra
$$\left[\overline{K}^{\mu\nu}, R_i\right] = -
\left(\overline{\sigma}^{\mu\nu}\right)_{ij} R_j\eqno(27)$$
$$\left\lbrace R_i, \tilde{R}_j \right\rbrace =
2\left(\overline{\sigma}^{\mu\nu}\right)_{ij}\overline{K}_{\mu\nu}
\eqno(28)$$
in place of (24-25).
We now turn to models invariant under the supersymmetry transformations
generated by $Q$ and $J^{\mu\nu}$.

\section{Supersymmetric Models for $AdS_3$ Supersymmetry}

It is apparent that with $Q_i$ defined by (19),
$$\left[Q_i, \Delta\right] = \left[ Q_i, A^2\right] = 0\eqno(29a,b)$$
where
$$\Delta = x^\mu \partial_\mu + \theta_i
\frac{\partial}{\partial\theta_i} = x^\mu \partial_\mu +
\tilde{\theta}_i\frac{\partial}{\partial\tilde{\theta}_i}\eqno(30)$$
$$A^2 = x^2 - \tilde{\theta}\theta .\eqno(31)$$

We also note that it is possible to define a superfield
$$F(x,\theta) = f_1(x) + \tilde{f_2}(x)\theta + f_3(x)
\tilde{\theta}\theta\eqno(32)$$
where $f_1$ and $f_3$ are scalars and $f_2$ is a Majorana spinor.  Under
a transformation generated by $Q$, we have
$$\delta x^\mu = \left[\tilde{\zeta} Q, x^\mu\right] =
\tilde{\zeta}\lambda^\mu \theta\eqno(33a)$$
$$\delta\tilde{\theta} = \left[\tilde{\zeta}Q, \tilde{\theta}\right] =
\tilde{\zeta} \lambda^\mu x_\mu\eqno(33b)$$
and consequently
$$\delta F(x,\theta) = \left(\tilde{\zeta}
\lambda^\mu\theta\right)\left(\partial_\mu \tilde{f}_2(x)\theta\right) +
\cal{O}(\theta).\eqno(34)$$
As a result, the change in a superfield induced by a supersymmetry
transformation is a surface term at order $\theta^2$.  Hence, if the
standard definition of Grassmann integration is adopted (with
normalization
$$\int d^2\theta\, \tilde{\theta}\theta = 1\eqno(34a)$$
or equivalently
$$\int d^2\theta\, \theta_i \theta_j =
\frac{1}{2}\left(\tau^2\right)_{ji}\;)\eqno(34b)$$
then a supersymmetric invariant action can be formed by integrating an
appropriate superfield (or product of superfields) over $x^\mu$ and
$\theta$.

For example, we might consider the action
$$S_0 = \int d^4x\, d^2\theta\, \delta\left(A^2 -
a^2\right)\Phi(x,\theta)\tilde{R} R\Phi(x,\theta).\eqno(35)$$
In order to define the superfield $\Phi$ off the surface of eq. (1), we
adopt the condition
$$\Delta \Phi = \omega\Phi \eqno(36)$$
where $\omega$ is a real constant.  This condition is supersymmetric
invariant by eq. (29a). A similar condition has been used by Dirac in
order to treat fields in four dimensional de Sitter space [6]. If $\Phi$
is expanded in terms of component fields
$$\Phi(x,\theta) = \phi(x) + \tilde{\lambda}(x)\theta + \frac{1}{2} F(x)
\tilde{\theta}\theta \eqno(37)$$
then the condition of eq. (36) leads to
$$x \cdot \partial \phi = \omega \phi\eqno(38a)$$
$$x \cdot \partial \psi = (\omega - 1)\psi\eqno(38b)$$
$$x \cdot \partial F = (\omega - 2)F.\eqno(38c)$$
It is possible to show from (26) that
$$\tilde{R} R = \tilde{\theta}\theta\left[
\frac{1}{x^2} \left(\frac{1}{2} J_{\mu\nu}J^{\mu\nu} + 2 x\cdot\partial
+ (x \cdot \partial)^2\right)\right] + 
\tilde{\theta}\frac{\partial}{\partial\tilde{\theta}} ( 4 + 2x \cdot
\partial)\nonumber$$
$$- 2x^2\,\frac{\partial}{\partial\theta}
\frac{\partial}{\partial\tilde{\theta}} - 2\tilde{\theta}
\sigma^{\mu\nu} \frac{\partial}{\partial\tilde{\theta}} J_{\mu\nu} -
2x\cdot\partial\eqno(39)$$
Also, we make use of the result
$$\int d^4x\, \delta \left( A^2 - a^2\right) G =
\int d^4x \left[ \delta\left(x^2 - a^2\right) - \tilde{\theta}
\theta\delta^\prime\left(x^2 - a^2\right)\right]G\nonumber$$
$$= \int d^4x \left[ \delta\left(x^2-a^2\right)\right]\left[
G + \frac{\tilde{\theta}\theta}{a^2} \left(G + \frac{1}{2} x \cdot
\partial G\right)\right].\eqno(40)$$
From eq. (40) it is apparent that the factor of $\delta(A^2 - a^2)$ in eq. (35) ensures that even though we are using embedding space coordinates in our superspace, we are considering only fields defined on the three dimensional subspace $AdS_3$ defined by eq. (1).
Together, (34), (37-40) reduce (35) to simply
$$\!\!\!\!\!S_0 = \int d^4x\, \delta\left(x^2 - a^2\right) \left[
\frac{1}{a^2} \phi\left(\frac{1}{2} J_{\mu\nu} J^{\mu\nu} - \omega^2\right)\phi
\right.\nonumber$$
$$\left. + \tilde{\psi}\left(\sigma^{\mu\nu} J_{\mu\nu} - 2\right) \psi
- \frac{a^2}{2} F^2 + \left(1 - \omega\right) \phi F\right].\eqno(41)$$
Supersymmetric invariant interactions can also be introduced, the simplest 
being of the form
$$S_I = \lambda_N \int d^4x\, d^2\theta\, \delta \left(A^2 - a^2\right)
\left[\Phi (x,\theta)\right]^N\;\;\;(N = 2, 3 \ldots ).\eqno(42)$$
Furthermore, in place of (35) it is possible to consider alternate actions possessing supersymmetry such as
$$S^\prime_0 = \int d^4x\, d^2\theta \,\delta \left(A^2 - a^2\right)
\left(\tilde{R}\Phi\right)\left(R\Phi\right).\eqno(43)$$
We now consider the representation of operators associated with
a supersymmetric extension of the $AdS_4$ algebra.

\section{$AdS_4$ Superspace}

In analogy with eq. (1), we take $AdS_4$ to be a four dimensional surface in a five dimensional embedding flat space
$$x^2 = \left(x^0\right)^2 - \left(x^1\right)^2 - \left(x^2\right)^2
- \left(x^3\right)^2 + \left(x^5\right)^2 = a^2.\eqno(44)$$
The associated Dirac matrices in the five dimensional embedding space are taken to be
$$\Gamma^0 = \left(
\begin{array}{cc}
0 & 1\\
1 & 0\end{array}\right)\;\;\;\;
\Gamma^5 = \left(
\begin{array}{cc}
-1 & 0\\
0 & 1\end{array}\right)\;\;\;\;
\Gamma^i = \left(
\begin{array}{cc}
0 & \tau^i\\
-\tau^i & 0\end{array}\right)\eqno(45)$$
so that a charge conjugation matrix $C$ satisfying $C\Gamma^AC^{-1} = 
\left(\Gamma^A\right)^T$ can be taken to be
$$C = i\Gamma^1\Gamma^3 = - \left(\begin{array}{cc}\tau^2 & 0\\
0 & \tau^2\end{array}\right) = 
C^{-1} = C^\dagger = -C^* = -C^T.\eqno(46)$$
Furthermore, if we let
$$\overline{Q} = Q^\dagger\left(-i\Gamma^0\Gamma^5\right)\eqno(47)$$
for a spinor $Q$ defined in the embedding space, and define charge conjugation by
$$Q_C = C\overline{Q}^T\eqno(48)$$
then
$$Q = (Q_C)_C\eqno(49)$$
so that $Q$ can be taken to be Majorana. Hence, if we define
$$\tilde{Q} = Q^TC\eqno(50)$$
a suitable superalgebra [2] associated with $AdS_4$ is
$$\left\lbrace Q, \tilde{Q} \right\rbrace = -2 \Sigma^{AB}
J_{AB}\eqno(51a)$$
$$\displaystyle{\left[J^{AB}, Q\right] = -\Sigma^{AB} Q}\eqno(51b)$$
$$\displaystyle{\left[J^{AB}, J^{CD}\right] = g^{AC} J^{BD} -
g^{BC}J^{AD}}\eqno(51c)$$
$$+ g^{BD}J^{AC} - g^{AD}J^{BC}\nonumber$$
with $\sum^{AB} = -\frac{1}{4}\left[\gamma^A, \gamma^B\right]$.  As in eqs. (24) and (25), we do not decompose the isometry generators on the embedding space, $J_{AB}$, into rotation and translation operators that are defined on the surface, as has been done in ref. [7].\\
(By using the Fierz identities
$$\displaystyle{\Sigma_{ij}^{AB} \Sigma_{AB\,k\ell} = -\frac{1}{2}\Sigma_{i\ell}^{AB}
\Sigma_{AB\,kj} - \frac{1}{4} \gamma_{i\ell}^A \gamma_{A\,kj}
-\frac{5}{4}\delta_{i\ell}\delta_{kj}}\eqno(52a)$$
$$\displaystyle{\gamma^A_{ij}\gamma_{A\;k\ell}
= -\frac{1}{2}\Sigma_{i\ell}^{AB}
\Sigma_{AB\,kj} - \frac{3}{4} \gamma_{i\ell}^A \gamma_{A\,kj}
+\frac{5}{4}\delta_{i\ell}\delta_{kj}}\eqno(52b)$$
$$\displaystyle{\delta_{ij} \delta_{k\ell} = -\frac{1}{2}\Sigma_{i\ell}^{AB}
\Sigma_{AB\,kj} + \frac{1}{4} \gamma_{i\ell}^A \gamma_{Akj}
+\frac{1}{4}\delta_{i\ell}\delta_{kj}}\eqno(52c)$$
all Jacobi identities associated with (51) can be shown to be
satisfied.)  Again, the Bosonic generator of translations has not
been included, in contrast to the superalgebras considered in [1].

A superspace is now introduced; it consists of the coordinates
$x^A$, a Majorana Fermion $\theta$ and an extra Bosonic variable
$\beta$ (whose significance is not apparent). It is an easy exercise
to show that a representation of the algebra of (51) is given by
$$Q = \left(\gamma_A\partial^A + \frac{\partial}{\partial\beta}\right)
\theta + \left(\gamma_Ax^A -
3\beta\right)\frac{\partial}{\partial\tilde{\theta}}\eqno(53a)$$
$$\tilde{Q} = -\tilde{\theta}
\left(\gamma_A\partial^A + \frac{\partial}{\partial\beta}\right)
+ \frac{\partial}{\partial\theta} \left(\gamma_{A}x^A -
3\beta\right)\eqno(53b)$$
$$\displaystyle{J_{AB} = -x_A\partial_B + x_B\partial_A +
\frac{\partial}{\partial\theta}
\Sigma_{AB} \theta}.\eqno(53c)$$
Two invariants, analogous to the two of eqs. (30) and (31) for $AdS_3$, are
$$\Delta = x^A\partial_A + \beta\frac{\partial}{\partial\beta} +
\tilde{\theta}\frac{\partial}{\partial\tilde{\theta}}\eqno(54a)$$
and 
$$A^2 = x^2 - 3\beta^2 - \tilde{\theta}\theta .\eqno(54b)$$

In our approach, in which the superspace uses $(3 + 2)$ dimensional Bosonic coordinates and Fermionic coordinates that are Majorana spinors in $(3 + 2)$ dimensions, it does not appear to be feasable to devise a realistic action for a superfield $\Phi (x,\theta)$. This is because there are now four independent components in the Majorana spinor $\theta$, so that in terms of component fields
$$\Phi (x,\theta) = f_1(x) + \tilde{f}_2 (x) \theta + f_3(x)(\tilde{\theta}\theta) + f_4^A(x)\tilde{\theta} \gamma_A\theta\nonumber$$
$$ + \tilde{f}_5(x)\theta(\tilde{\theta}\theta) + f_6(x)(\tilde{\theta} \theta)^2\; .\eqno(55)$$
($f_1, f_3, f_6$ - scalars, $f_4^A$ - vector, $\tilde{f}_2, \tilde{f}_5$ - Majorana spinors)\\
With $\int d\theta_i\theta_j = \delta_{ij}$, it follows that in this formulation, actions of the form $\int d^4x \int d^4\theta \Phi (x,\theta)D^2 \Phi(x,\theta)$ would not be realistic as, for example, spinor fields would have kinetic terms involving second derivatives. This is essentially because the Grassmann coordinate $\theta$ has four independent components, being a Majorana spinor in $(3 + 2)$ dimensions. 
As will be argued below, it is possible to use the rotationally invariant projection operators [8] $\gamma_{\pm} = \frac{1}{2} (1 \pm \gamma \cdot x/\sqrt{a^2})$ to reduce the number of independent components of $\theta$ from four to two, thereby making it feasable to formulate a realistic model in terms of $AdS_4$ superfields in our embedding superspace.

\section{Discussion}

We have succeeded in formulating representations of the supersymmetry algebras associated with $AdS_3$ and $AdS_4$ using a superspace involving embedding space variables. This approach closely resembles that used in conjunction with the symmetry associated with $S_2$ and $AdS_2$ in ref. [2].

The approach we have employed using an embedding space is quite distinct from the standard one outlined in ref. [7]. There, supersymmetric models on $AdS_4$ space are constructed using a superspace in which the Bosonic coordinates are variables characterizing the curved surface and Fermionic coordinates are chiral spinors defined in this four dimensional space.

Also, in ref. [7] the isometry generators of the embedding space have been decomposed into ``translation'' ($R$) and ``rotation'' ($M$) operators defined on the surface with $[R,R] \sim M$. Furthermore, in ref. [7] spinorial generators the supersymmetry algebra associated with $AdS_4$ are spinors in the space itself rather than the embedding space.

To show how our approach to $AdS_4$  is related to that of ref.[7], we begin by considering the vectors $x^A$ and $y^A$ in the five dimensional embedding space of $AdS_4$. Following ref. [9], if $h^A$ is a unit vector in five dimensions, the conformal transformation
$$ x^A = ah^A + 2 \frac{y^A - h^A y^2/a}{1-2y \cdot h/a + y^2/a^2}\eqno(56)$$
maps the plane $y \cdot h = 0$ onto the surface $x^2 = a^2$ of eq. (44). In this paper we have been using the coordinates $x^A$; in ref. [7] coordinates $y^A$ on the plane $y \cdot h = 0$ have been employed with $h^A = (0,0,0,0,1)$. Furthermore, the number of independent components of the Grassmann coordinates $\theta$ contributing to a superfield $\Phi (x,\theta)$ have been reduced using the projection operators $\frac{1}{2} (1 \pm \gamma \cdot h)$. It is this reduction that leads to realistic actions after integrating over Grassmann coordinates in the super field actions in ref. [7]. This sort of projection is contingent to being in four dimensions.  Kinetic terms for the scalar and spinor fields in the models of ref. [7] are easily shown, using the techniques of ref. [9], to reduce to $\phi(L^2 - 2)\phi$ and $\overline{\psi}(\Sigma^{AB}L_{AB})\psi$ respectively when using the coordinates of the embedding space.

Similarly the review of ref. [10] handles SUSY in AdS spaces in a way that is distinct from the treatment presented in this paper. In ref. [10], the spinors arising in $d$-dimensional AdS spaces are $d$-dimensional spinors, while we have used spinors defined in the $d + 1$ dimensional embedding space that transform under $SO(d-1,2)$. Furthermore, superfield techniques are not employed in ref. [10], while we have considered irreducible representations of the $AdS_3$ SUSY algebra and generated a suitable off-shell action by working with superfields defined in a superspace whose Bosonic part is a $d$-dimensional surface embedded in $d + 1$ dimensions. In addition, the masses for the fields appearing in eq. (41) that we have derived using superfields are quite distinct from those of ref. [10] where on-shell closure of the postulated SUSY transformations of the fields is used to determine the masses of the component fields.

It is not clear if the approach to using superfields to analyze SUSY in $AdS_2$ and $AdS_3$ presented in ref. [2] and in this paper can be extended to treat SUSY models in higher dimensional spaces. In particular, for $AdS_5$ this would entail employing six dimensional spinors transforming under $SO(4,2)$. These spinors have eight complex components, and no irreducible  representation of $SO(4,2)$ consists of spinor with only two independent components--which, as is discussed above, is a necessary condition for having a viable superfield model. We do anticipate though that the superfield models introduced in conjunction with $AdS_2$ and $AdS_3$ will be of interest in the context of AdS/CFT correspondence as brane solutions to higher dimensional supergravity models containing $AdS_2$ as a subspace are possible [11].

We anticipate that it will prove possible to analyze the quantum properties of our $AdS_3$ model.

\section{Acknowledgements}

D.G.C. McKeon would like to thank IFM at the Universidad de Michoacana
de San Nicolas de Hidalgo where much of this work was done and A. Buchel for conversations. Funding was
provided by IFM and 
NSERC. R. and D. MacKenzie
gave helpful advice.

\eject

\end{document}